\documentclass[iop]{emulateapj}
\usepackage{natbib} 
\usepackage{psfig} 
\shorttitle{X-ray Variability of NGC~3227}
\shortauthors{P. Ar\'evalo \& A. Markowitz}

\def\mnras{MNRAS}
\def\apj{ApJ}
\def\aap{A\&A}
\def\apjl{ApJL}
\def\nat{Nature} 
\def\apjs{ApJS}

\def\xmm{{\sl XMM-Newton}}
\def\rms{$\sigma_{\rm rms}$}

\def\rxte{{\sl RXTE}}
\def\ngc{{NGC~3227}}

\def\nh{$N_{\rm H}$}
\def\Msun{M$_\odot$}

\begin{document}

\title{Deconvolving X-ray Spectral Variability Components in the Seyfert 1.5 NGC~3227}  
\author{P.\ Ar\'evalo \email{parevalo@astro.puc.cl}}
\affil{Pontificia Universidad Cat\'olica de Chile, Instituto de Astrof\'isica, Casilla 306, Santiago 22, Chile}
\author{A.\ Markowitz}
\affil{Center for Astrophysics and Space Sciences, University of California, San Diego, Mail Code 0424,  La Jolla, CA 92093-0424, USA}

\begin{abstract}
We present the variability analysis of a 100 ks \xmm\ observation of the Seyfert 1.5 active galaxy NGC 3227.  The observation found NGC 3227 in a period where its hard power-law component displayed remarkably little long-term variability. This lucky event allows us to observe clearly a soft spectral component undergoing a large-amplitude but slow flux variation.  Using combined spectral and timing analysis we isolate two independent variable continuum components and characterize their behavior as a function of timescale.  Rapid and coherent variations throughout the 0.2-10 keV band reveal a spectrally hard (photon index $\Gamma \sim 1.7-1.8$) power law, dominating the observed variability on timescales of 30 ks and shorter.  Another component produces coherent fluctuations in 0.2--2 keV range and is much softer ($\Gamma \sim 3$); it dominates the observed variability on timescales greater than 30 ks.  Both components are viewed through the same absorbers identified in the time-averaged spectrum.  

The combined spectral and timing analysis breaks the degeneracy between models for the soft excess: it is consistent with a power-law or thermal Comptonized component, but not with a blackbody or an ionized reflection component.  We demonstrate that the rapid variability in NGC 3227 is intrinsic to continuum-emitting components and is not an effect of variable  absorption.  

\end{abstract}


\section{Introduction}       

There is general consensus that the X-ray continuum emission in
Seyfert Active Galactic Nuclei (AGN) and black hole X-ray Binaries (BH XRBs)
is mainly produced by Compton
up-scattering of lower energy photons in a hot ($\sim10^9$ K) corona; 
these seed photons are likely produced
thermally by the accretion disc which feeds the supermassive black hole \citep{haardt91}.
The resulting Comptonized spectrum is a power
law with high- and low- energy cut-offs constrained, respectively, by 
the temperature of the Comptonizing medium and by the
energies of the seed photon population \citep{sunyaev80}.
Many Seyfert AGN spectra however exhibit
an additional spectral component in the X-ray regime that manifests itself as an excess
of emission over the power-law continuum below approximately 1 keV
\citep{turner89}. The interpretation of the so-called soft excess is still a matter of
debate, with models such as thermal Comptonization (Mehdipour et
al.\ 2011), relativistically-blurred reflection off ionized gas
\citep[e.g.][]{RossFabian2005, Crummy2006}, or absorption \citep{GierlinskiDone2004} frequently invoked to model the soft
emission.  

In this paper we explore the X-ray continuum emission components
originating in the accretion disk/Comptonization corona of the Seyfert 1.5 AGN NGC~3227 
($z$=0.00386; \citealt{deVaucouleurs1991}); we
combine energy spectral fitting with
variability analysis to deconvolve the X-ray spectral components that vary
coherently.

\citet[hereafter M09]{markowitz_3227} modeled in detail the X-ray
energy spectrum of NGC~3227 using a long look (100 ks continuous)
observation with the \textit{XMM-Newton} observatory.  They modeled a
continuum composed of a hard ($\Gamma \sim 1.6$) power law dominating
the 1--10 keV band and a moderate soft excess below 1 keV.  The soft
excess was modeled approximately equally well by a simple steep
($\Gamma \sim 3.4$) power law, a Comptonized spectrum resulting from
scattering off optically-thick ($\tau \sim 24$) gas with electron
temperature $\sim$0.35 keV, or a phenomenological blackbody
component with temperature $k_{\rm B}T\sim 0.08$ keV.  A further complication is
the presence of at least three zones of absorbing gas lying along the sight.
These include a layer of cold absorption with column density $N_{\rm
  H, cold} \sim 5-9 \times 10^{20}$ cm$^{-2}$ plus two ionized absorbers,
one a lowly-ionized (log($\xi$, erg cm s$^{-1}$) $\sim
1.2-1.4$\footnote{$\xi \equiv L_{\rm ion}/(nr^2)$, where $L_{\rm ion}$
  is the luminosity of the ionizing continuum, $n$ is the number
  density of the gas, and $r$ is the distance from the source of the
  ionizing continuum to the gas cloud. All values of $\xi$ in this
  paper are in quoted in units of erg cm s$^{-1}$.}) absorber
outflowing at $\sim$400 km s$^{-1}$ relative to the systemic velocity, the other
more highly-ionized (log($\xi$) $\sim$2.9) and outflowing at
$\sim$2100 km s$^{-1}$ relative to systemic; both ionized zones had
column densities $\sim 1-2 \times 10^{21}$ cm$^{-2}$.  A way to tackle
the model degeneracy is to combine the spectral analysis with analysis
of the source variability as a function of energy band. The goal is to
shed light on the possible connections between the continuum emission
components, and importantly, on the (not yet fully-understood)
emission and variability mechanisms.

The rest of this paper is organized as follows: $\S$\ref{observations} summarizes the
observations and light curve extraction. In $\S$\ref{pds}, we use the new data to 
re-measure the 2--10 keV power density spectrum (PDS) for this source,
previously published by \citet[hereafter UM05]{uttleymchardy05}
In $\S$\ref{CCF}, we 
compare variability trends in soft and hard bands
over various timescales and measure interband coherence functions.
We examine the energy spectra of the variable components in $\S$\ref{rmsspec} and \ref{fits}.
We interpret the spectral components in the context of previous observations and comparison to other sources in $\S$\ref{interpretation}. A summary of our main results is given in $\S$\ref{conclusion}.

\section{Observations and Light Curve Extraction}
\label{observations}
We used data from a \xmm\ full-orbit observation of \ngc\ taken on
2006 December 3--4 (obsID 0400270101). These data have been previously presented in
M09, where the analysis was focused on energy spectral
fitting. In this paper we consider data only from the EPIC pn detector and
the focus of the analysis is on the variability properties of the
light curves.

The EPIC pn camera \citep{struder} was operated in Large
Window mode, using the medium filter. Single and double events were
extracted from a circular region of $40\arcsec$ radius centered on
NGC~3227 and background photons were extracted from a source-free
region of the equal area on the same chip. Only quality flag=0 photons
were selected for both source and background. As noted by M09,
pile-up on these pn data is negligible. The
background-subtracted, 0.2--10 keV average count rate in the pn
observation was 11.7 count s$^{-1}$.

\begin{figure}
\psfig{file=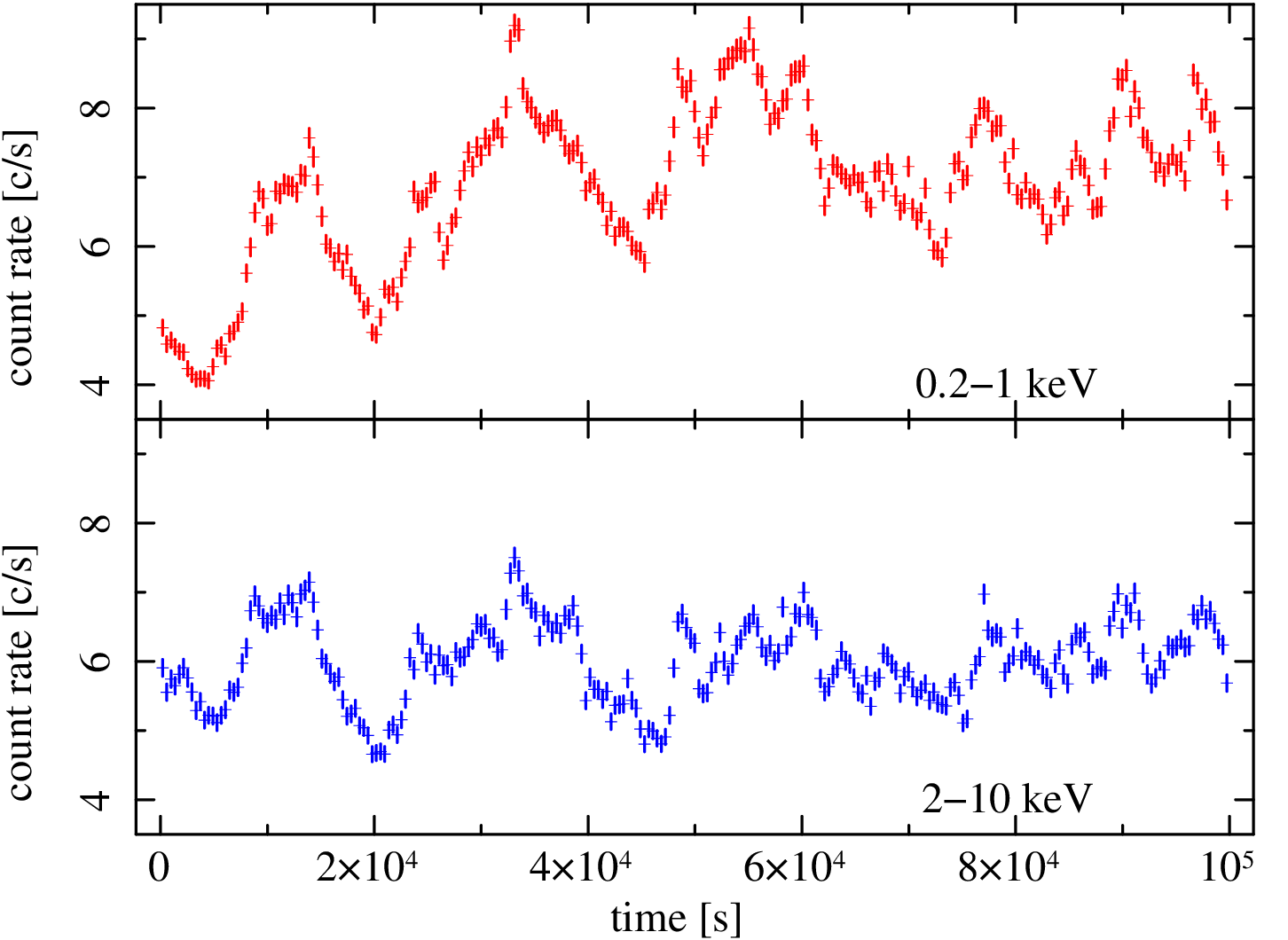,angle=270,width=8.0cm,height=5.5cm}
\caption{\label{fig:lcs} \xmm\ EPIC light curves of NGC~3227 in soft (0.2--1 keV) and hard (2--10 keV) energy bands. 
The soft band displays higher amplitude of rapid fluctuations and a long term trend not visible in the hard band.}
\end{figure}

Light curves were constructed in several energy bands as described in
the analysis below.  The count rates were corrected for dead time in
each time bin, according to the good time interval file of the
corresponding chip.  Light curves for the soft (0.2--1 keV) and hard (2--10 keV) energy bands 
are displayed in Fig.\ \ref{fig:lcs}.

\section{Power Density Spectrum Analysis}
\label{pds}
The broadband timing properties of \ngc\ have been studied previously by 
UM05, who used data from four multi-timescale monitoring campaigns 
from the \textit{Rossi X-ray Timing Explorer} (\textit{RXTE}), plus a long-look from \textit{EXOSAT}.
These campaigns consisted of regularly-spaced $\sim$1-ks snapshots,
with sampling patterns and durations chosen such that the campaigns
probe complementary ranges in temporal frequency space. This yielded a 2--10 keV 
power density spectrum (PDS) spanning from
$\sim 7 \times 10^{-9}$ to $\sim 3 \times 10^{-3}$ Hz.
UM05 fit a bending
power-law model, 
$P$($f$) = $A f^{\alpha_{\rm L}}$ [  1 + ($f/f_{\rm b}$)$^{(\alpha_{\rm L} - \alpha_{\rm H})}$]$^{-1}$, where
$f_{\rm b}$ is the bend frequency, $\alpha_{\rm L}$ and $\alpha_{\rm H}$ are the power-law slopes
below and above $f_{\rm b}$, respectively, and $A$ is a normalization factor.
They found a bend at $f_{\rm b}=2.6^{+6.1}_{-1.8} \times 10^{-5}$ Hz, with
best-fit values of $\alpha_{\rm L} = -1.0 \pm 0.1$ 
and $\alpha_{\rm H} < -2.0$.
Strong energy dependence was evident from the
normalizations of the sub-band PDSs, where the lower energy band (3--5
keV) varied more strongly than the 7--15 keV band. 

\begin{figure}
\psfig{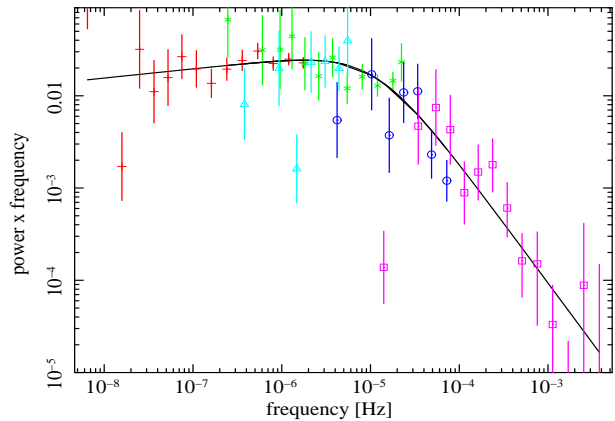}
\caption{\label{fig:pds} PDS data are denoted by points with error bars.  
The PDSs are unfolded and plotted in ``model space'', i.e., plotted relative to the best-fit 
bending power-law model (solid line) and with the effects of Poisson noise and
PDS measurement distortion effects removed. 
The red, green, cyan, and blue points denote the PDS segments derived from the ``long-term'', ``six-hourly'', ``daily''
and ``long-look'' \textit{RXTE} campaigns, respectively (see UM05 for details of observations for each campaign).
The \xmm\ data are represented by the pink squares, covering the high frequency region. 
Notice the unusually low variability power of the lowest frequency \xmm\ data point.}
\end{figure}

We used the EPIC light curves to both recalculate the high-frequency section of the PDS and 
investigate the energy
dependence of variability amplitude extending to below 2 keV.
For the former, we used the same \textit{RXTE} Proportional Counter Array (PCA) data as in UM05, and extracted 
2--10 keV lightcurves using standard extraction techniques (see e.g., \citealt{markowitz03} or UM05)
and the ``pca\_bkgd\_cmfaintl7\_eMv20051128.mdl'' PCA background files.
We combined the PDS derived from the 2--10 keV \xmm\ light curve with those derived from the  
``long-term'', ``six hourly,'' ``daily,'' and  ``long-look'' \textit{RXTE} campaigns described in UM05.
The only difference is that the ``long-term'' campaign published in UM05 contained monitoring
data from 1999 Jan 2 to 2005 Feb 24, but here we used monitoring up until 2005 Dec 4.
We binned the periodograms  
using a minimum of 2 points per bin and a frequency binning of $\Delta f=1.5 f$. 
We used the PSRESP method \citep{psresp} to model the   
PDS with a bending power law; the observed PDS and best-fit model are
shown in Fig.~\ref{fig:pds}. This unfolded power spectrum is 
Poisson Noise-subtracted and the distortion effects
caused by aliasing and red-noise leak have been removed. 

Our best-fit bend frequency is $1.15^{+0.76}_{-0.81}\times 10^{-5}$
Hz, consistent with UM05. The high-frequency slope is
$\alpha_{\rm H}<-2.0$, where errors quoted 
correspond to 68\% 
bounds. For the low-frequency slope, the best-fit value was
$\alpha_{\rm L}=-0.9$, with uncertainties pegging at the boundaries of the 
range over which we tested $\alpha_{\rm L}$, --0.8 to --1.0 
(the current PDS does not offer any ability for us to constrain the
low-frequency shape of the PDS any better than UM05, so we restricted 
$\alpha_{\rm L}$ to encompass the best-fit values from UM05).
The \xmm\ PDS data, plotted in pink in Fig.~\ref{fig:pds},
show that for this \xmm\ light curve the variability power at
the longest timescales probed ($\sim100$ ks) is unusually low. Although this is not
strange given the stochastic red noise behavior of AGN variability \citep{Vaughan2003}, 
it does give us the rare opportunity to study in detail the variability of components other
than the hard power law.  In the following sections we will attempt to
disentangle the spectral components of the variability.

{ The power spectra of the \xmm\ light curves were computed with the
Mexican-hat procedure described in \citet{rmsk}. This method convolves the lightcurves with a band-pass filter and then computes the variance of the filtered lightcurve as a measure of the variability power as a function of variability timescale. The power spectrum was calculated for three energy
bands, soft (0.2--1 keV), medium (1--2 keV) and hard (2--10 keV). }
Figure~\ref{pds} shows these power spectra for each band in black filled
circles, blue crosses and open magenta circles, respectively. The
\xmm\ power spectra taken on their own appear to have higher frequency
breaks (near $\sim 5\times 10^{-5}$ Hz) than the complete data set. This is probably an artifact of the
unusually small amplitude of variability on the longest timescales of
the \xmm\ observation. 
At frequencies $\lesssim 5\times 10^{-5}$ Hz, the amount of power in the three bands diverges sharply.
We explore the energy dependence of the variability in further detail in the following sections.

\begin{figure}
\psfig{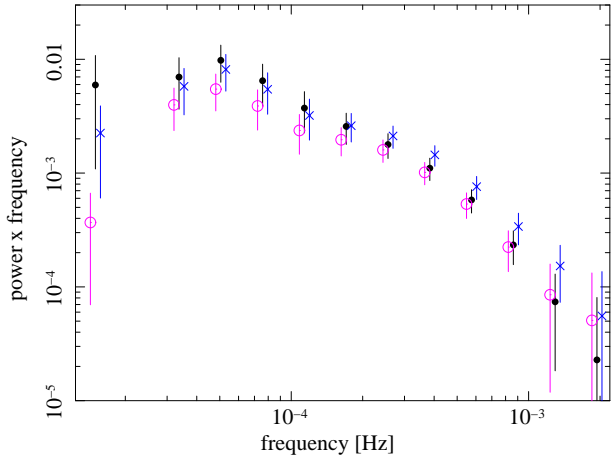}
\caption{\label{pds} Power spectra calculated using only the \xmm\ lightcurves in the soft band, 
0.2--1 keV (black, filled circles), medium band, 1--2 keV (blue crosses) and hard band, 2--10 keV band 
(magenta open circles). The 1--2 keV points have been shifted slightly to lower 
frequencies and the 2--10 keV point to higher frequencies to avoid overlapping error bars. }
\end{figure}

\section{Interband Cross-correlation}
\label{CCF}

As demonstrated in the previous section, the \xmm\ light curves in
different energy bands display qualitatively similar variability patterns but with
different amplitudes on different timescales. In this section, we will investigate further the dependence of variability amplitudes on time scale and energy band and we will then investigate the degree of correlation between the variations in different energy bands.

We constructed light curves from the pn data for 22 energy bands in
the 0.2--10 keV range. The energy resolution was chosen to allow 
similar number counts in each band, taking care to include integer
numbers of pn energy channels. For the spectral fitting, the pn
response matrix was re-binned using the same energy resolution of the
light curves by using the task \textsc{rbnrmf}.

We isolated variability trends over various timescales in each light
curve by applying a Mexican-hat type filter: in this method, described
in \citet{rmsk}, one convolves each light curve with two
Gaussian profiles, of slightly different widths, and takes the
difference of the convolved light curves. This convolution and difference procedure is equivalent to a multiplication of the power spectrum by a filter which peaks at a frequency equal to $0.225/\sigma$. This method thus isolates
variations on timescales of $\sigma/0.225$ where $\sigma$ is the average
width of the Gaussian filters, effectively removing variations on
shorter and longer timescales. Varying the value of $\sigma$ we can isolate the variability pattern and amplitude as a function of filter timescale or equivalently, frequency, to construct filtered light curves and power spectra. This is particularly useful to remove
effects of red-noise leak, which would hamper a Fourier analysis
alternative to decompose the variability patterns at different
timescales, especially for the low energy cases which show a strong
trend on timescales longer than the observation. For the analysis below we used pairs of Gaussian kernels separate in width by a factor of 1.01 but the results are quite insensitive to the choice of width separation as explained in \citet{rmsk}. 

\begin{figure}
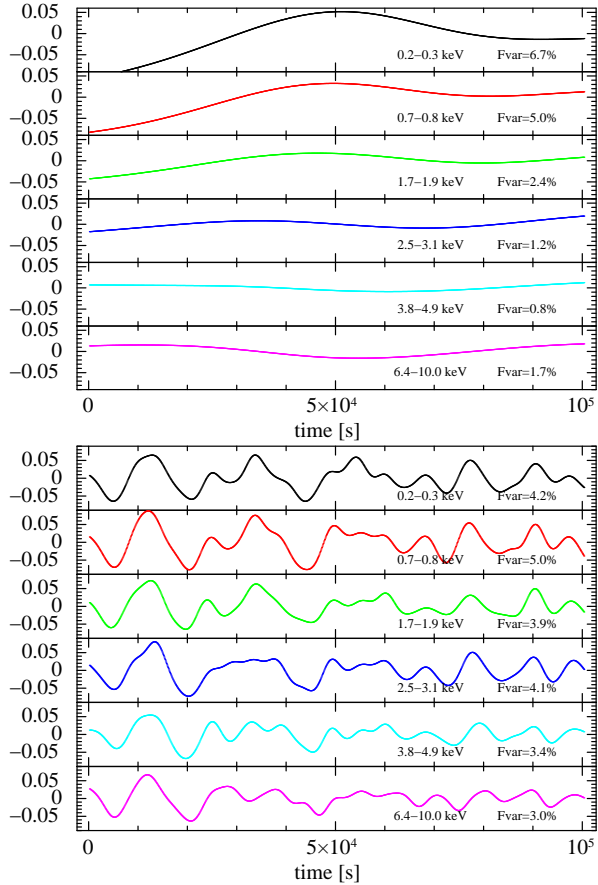

\psfig{file=filteredlcs_1_7_15_18_20_22.ps,angle=270,width=8.0cm}
\psfig{file=filteredlcs_1_7_15_18_20_22_k3.ps,angle=270,width=8.0cm}
\caption{\label{fig:filtered_lcs} \textit{Top panel:} Light curves in various
energy bands filtered to show only long-term (100 ks) fluctuations; 
the variations change shape at energies around 2 keV. 
\textit{Bottom panel:} light curves in the same energy bands, now 
filtered for shorter timescale fluctuations (5.9 ks). 
The variations remain coherent throughout the energy range.}
\end{figure}

We filtered the light curves with this procedure in two different timescales, 100 ks and 10 ks. 
In the top panel of Fig.~\ref{fig:filtered_lcs}, we plot the 100 ks trends 
for a range of selected energy bands. The
softest energy bands have the largest amplitudes of
fluctuations. Moreover, the pattern of variability changes above
energies of 2 keV, providing evidence for two or more distinct
spectral components. In the bottom panel, the rapid, 10 ks fluctuations
in the same energy bands are shown. For these shorter timescales, all
the energy bands from 0.2 to 10 keV show consistent fluctuation patterns.

\begin{figure}
\psfig{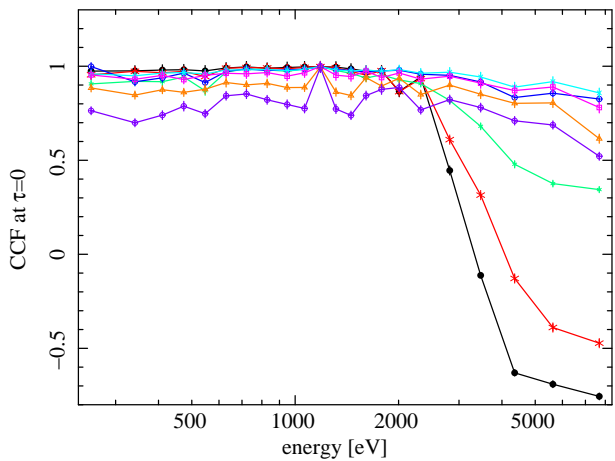}
\caption{\label{fig:dcf}
Coherence between the 1.1--1.25 keV light curve and variations in all other energy bands for fluctuations in eight
timescales. Black, red, green, blue, cyan, magenta, orange and purple, 
probe variations on timescales of 100.3, 66.8, 44.6, 29.7, 19.8, 13.2, 8.8, and 5.9 ks, respectively. The CCF value at 0 time lag is plotted in the y-axis. 
}
\end{figure}

Next, we quantified the coherence between energy bands by calculating the cross
correlation function (CCF) between the filtered light curves for 100.3, 66.8, 44.6, 29.7, 19.8, 13.2, 8.8, and 5.9 ks, the eight timescales used to
generate \rms\ spectra (described below). We chose an intermediate
energy band (1.1--1.25 keV) as reference and computed the CCF between this band and
each of the other bands. The CCF peak values as a function of energy band
are plotted in Fig.~\ref{fig:dcf}.

The slower variations (timescales longer than $\sim 40$ ks) 
are fully coherent with the reference band up to
energies of 2 keV. Above this energy, the light curves lose coherence
with the variations in the reference band, approaching an
anti-correlated trend with a central CCF value of almost --0.8. The
light curves are too short to establish whether this anti-correlation
is significant or just a chance offset of half a cycle between
uncorrelated light curves. It is clear, however, that for these long
term trends, the variations are synchronized for all bands between 0.2
and 2 keV and above this energy the variability pattern changes.  In
contrast, rapid fluctuations (timescales $\leq 30$ ks) are highly coherent throughout the
energy range probed.  This result indicates that one spectral
component covers the 0.2--10 keV energy range and is varying
coherently on these short timescales.

{ The rapid fluctuations are quasi-simultaneous across the whole energy range but there are small delays between different energy bands, as can be seen in the bottom panel of Fig. \ref{fig:filtered_lcs}. The peak of the CCF function shifts slightly with increasing energy separation. For light curves filtered at timescales around 19.8 ks (corresponding to the cyan line in Fig. \ref{fig:dcf} and \rms\ spectrum \#5 below), the CCF peaks at a maximum lag of 500 s between the reference band and the highest energy band. This time lag corresponds to about 2\% of the variability timescale.  Similar time lags, where the harder bands lag the softer ones, are commonly seen in the powerlaw component of Seyfert galaxies \citep[e.g.][]{papadakis_lags,Vaughan2003,McHardy4051,markowitz05,arevalo06,arevalomkn335,walton13b}.}

\section{Spectra of the Variable Components}
\label{rmsspec}
The coherence analysis shows that the difference in the soft and hard
light curves in Fig.~\ref{fig:filtered_lcs} arises from at least two spectral
components varying on different timescales. Here we will explore in
detail this energy dependence by determining the spectral shape of the
variable components. 

{  The spectral shape of the variable components can be studied using the \rms\ spectrum \citep{revnivtsev99} where the flux at each energy bin is replaced by the amplitude of \rms\ fluctuations in the lightcurve of the corresponding energy band. We note that the stochastic nature of the X-ray lightcurves means that the power of any Fourier mode varies greatly from one observation to another around the underlying mean value. High-frequency modes can normally be binned together so that many modes fall in the same bin and the average value is always closer to the underlying power spectrum. Low-frequency modes are more sparse and only few modes can be binned together so the scatter between low frequency powerspectral points of different observations is large.  The \rms\ spectrum is a collection of powerspectral estimates at a fixed frequency for different energy bands. If the energy bands are correlated, then a simultaneous observation will catch them all in the same fluctuation, so that if in our particular observation the power-spectrum estimate is higher/lower than the average it will be so for all the energy bins. Therefore, for energy bands that vary in a correlated way, even a single Fourier mode can be used to calculate the \emph{shape} of the \rms\ spectrum, even if the \emph{normalization} is uncertain due to the stochastic nature of the light curves. Another consequence of the simultaneity of the observations is that the point-to-point scatter between \rms\ spectrum bins is produced by the uncorrelated variability only, i.e. the Poisson noise and additional small amplitude variable components that might affect one band more than others, not by the stochastic nature of the intrinsic variability.  } 

We constructed \rms\ spectra by calculating \rms\ variability amplitudes for
band-pass-filtered lightcurves as a function of energy. The result is the
spectral shape of all components that vary on that
particular timescale, i.e. the \rms\ spectrum.  As a reminder, the
energy spectrum describes the distribution of emitted/absorbed photons
as a function of energy regardless of variability or lack thereof; the
\rms\ spectrum, meanwhile, isolates the contributions from only those
spectral components that are variable over a given timescale. Any constant additive components are automatically removed from the \rms\ spectra while constant multiplicative components appear with the same shape as in a flux spectrum. Variable multiplicative components, such as variable absorption, produce peaks in the \rms\ spectrum at energies where they produce the largest changes, normally where the absorption is strongest and dips would appear in the flux spectrum. One can
calculate and fit \rms\ spectra and compare to those components
modeled in the time-averaged energy spectrum to determine which
component(s) are relevant to explaining observed variability for a
given timescale/temporal frequency range.

We thus repeated the filtering procedure with different values of $\sigma$
to track the evolution of the \rms\ spectrum with variability
timescale.  We probe in total eight timescales ranging from  
100 ks (henceforth referred to as \#1) down to 5.9 ks (\#8). 
{ The power of the Poisson noise in each light curve is estimated from the light curve error bars, which incorporate the uncertainties due to counting noise of the photons in the source and background time bins. For the power spectrum normalization used here, the Poisson noise power PN$=1/(\bar{\rm{err}}^2\times dt)$, where  $dt$ is the time bin in s and $\bar{\rm err}$ is the average error of the lightcurve bins. The Poisson noise is white noise and therefore its contribution is the same at every temporal frequency. This estimate of the power due to Poisson noise is thus subtracted from the
variance measurement at each timescale and the process is repeated for each energy band. We note that since all the energy bands have similar count rates their Poisson noise contributions are also similar. The resulting
\rms\ spectra denote the intrinsic variability of each band, in units of
count s$^{-1}$ keV$^{-1}$. }Higher frequency \rms\ spectra have lower intrinsic variability power so the Poisson noise contribution is relatively stronger. Therefore, the subtraction of Poisson noise power produces a larger uncertainty in higher frequency \rms\ spectra and for this reason they have larger error bars.

As can be seen in Fig. \ref{fig:dcf}, different energy bands are highly correlated but the correlation is not perfect. This is partly due to the uncorrelated contribution of Poisson noise to each band and partly an effect of minor spectral component displaying uncorrelated variations. These additional sources of variability introduce power to some bins in the \rms\ spectrum that add to the \rms\ power of the fully coherent fluctuations of a single spectral component. Since it is not clear \textit{a priori} which energy bins are carrying the uncorrelated source of variability, it is not possible to subtract this contribution, but it is possible to incorporate its average effect on the errors.  We used the drop in coherence between energy bands to estimate the contribution of uncorrelated fluctuations (Poisson and intrinsic) to the variability at each timescale. We then introduced this contribution as uncorrelated error in the corresponding \rms\ spectra, knowing that the true correlated \rms\ power must be within this value of the measured power.

\begin{figure}
\psfig{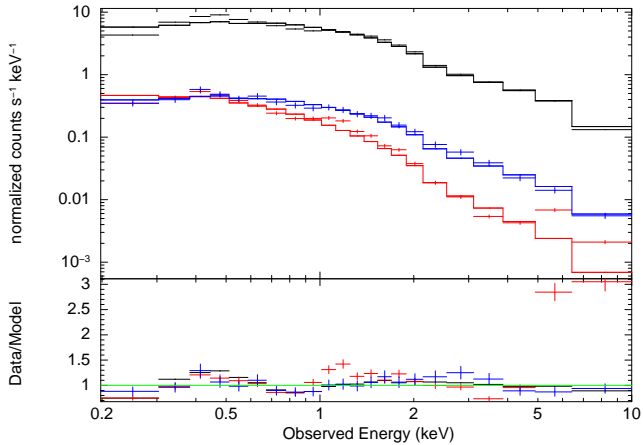}
\caption{\label{fig:rms_spec_first} 
$\sigma_{\rm rms}$ variability spectra for low-frequency 
fluctuations ($1\times10^{-5}$ Hz; spectrum \#1) in red 
and high-frequency fluctuations ($5\times10^{-5}$ Hz; spectrum \#5) 
in blue, together with the time-averaged spectrum in black. The spectra are 
fitted with a simple power law affected only by neutral absorption due 
to the Galactic column. Spectrum \#1 is fit only over 0.2--4 keV because above this energy the fluctuations are incoherent with those below about 2 keV.}
\end{figure}

{ Two of the \rms\ spectra, \#1 and \#5, are plotted in Fig.~\ref{fig:rms_spec_first},
to illustrate the change in spectral shape with timescale. Spectrum  \#1 correspond to the slowest fluctuations, where the difference between soft and hard light curves is strongest. Spectrum  \#5 is characteristic of the more rapid fluctuations --- at this frequency and above the fluctuations of the entire energy range are well correlated.} Also plotted is the time-averaged spectrum, degraded to match the
resolution of the \rms\ spectra. 

For all spectral fitting, we used \textsc{xspec} version 12.8.0;
uncertainties on best-fit model parameters
correspond to $\Delta\chi^2=+2.71$ (90\% confidence for one parameter)
unless otherwise noted.
As a preliminary step, we fit the three spectra with a crude model consisting solely of a simple power law.
This model and all models discussed herein contain 
neutral absorption due to the Galactic column, $1.98 \times 10^{20}$ cm$^{-2}$  \citep{Kalberla2005},
modeled with \textsc{phabs}.
Data/model residuals are plotted in the lower panel of Fig.~\ref{fig:rms_spec_first}

Spectrum \#1 was fit only in the 0.2--4 keV range, and there is a strong
increase in variability power above the extrapolation of the power law. 
This does not mean, however, that the soft spectral component has an upturn at
high energies. From the coherence analysis it is clear that above $\sim$2
keV, a different spectral component begins to dominate the variability,
so that by $\sim$5 keV the fluctuations are uncorrelated to the soft
component. 

This behavior is consistent with the two-component interpretation, where the components are varying incoherently on long timescales with different amplitudes. At low energies the variability is dominated by the large amplitude of the soft
component variations, but in higher energy bands this soft component
is weak, so the low-amplitude variability of the hard component takes
over. Near 2 keV, where the coherence changes
from 1 to --0.8, the variations interfere, since the value of the coherence
coefficient of --0.8 indicates that the fluctuations of both components are
almost exactly out of phase. Therefore, the variance in these
intermediate bands that have similar amplitude contributions from soft
and hard components is reduced, producing an artificial dip in the
\rms\ spectrum. The coherent variability from 0.2 to 2 keV however
ensures that the \rms\ spectrum in this energy range can be
interpreted as a single component that truly reflects the shape of the
emitted spectrum.  Above this energy we can expect contributions from
two or more spectral components whose variations interfere, distorting
the shape of the \rms\ spectrum. Below, we thus fit spectra \#1--\#3 only up to 3 keV;
a simple power-law fit yields $\Gamma \sim 2.33$.
As we will demonstrate below, spectrum \#1 is fairly representative of spectra \#1--\#3,

Spectrum \#5, meanwhile, is much flatter, with $\Gamma \sim 1.57$.
Since fluctuations in this timescale are highly
coherent throughout the entire energy range, the \rms\ spectrum can
be more simply interpreted as a single physical spectral component.
As will be demonstrated below, spectrum \#5 is representative of spectra \#4--\#8.

\section{Model Fits to the \rms\ Spectra}
\label{fits}
The continuum of the time-averaged spectrum is also modified by as-yet 
unmodeled line of sight absorption (M09), as evidenced by the dip
near 0.8 keV.  Models consisting of a simple power law are poor fits,
with $\chi^2/dof$=201/16=12.5 and 29.5/20=1.5 for spectra \#1 and \#5,
respectively.  The data/model residuals for both \rms\ spectra are
similar: there are $\pm30-40\%$ deviations, significant curvature below
0.4 keV, and a strong dip near 0.7--0.9 keV, the energies most
strongly affected by the low-ionization (log($\xi$)$\sim1.2-1.5$)
absorber modeled in the EPIC and Reflection Grating Spectrometer (RGS)
energy spectra by M09.  We now demonstrate that the \rms\ spectra are
consistent with being modified by the same absorbers while
simultaneously exploring mechanisms to explain their form.

\subsection{Low-frequency \rms\ spectra}
\begin{figure}
\psfig{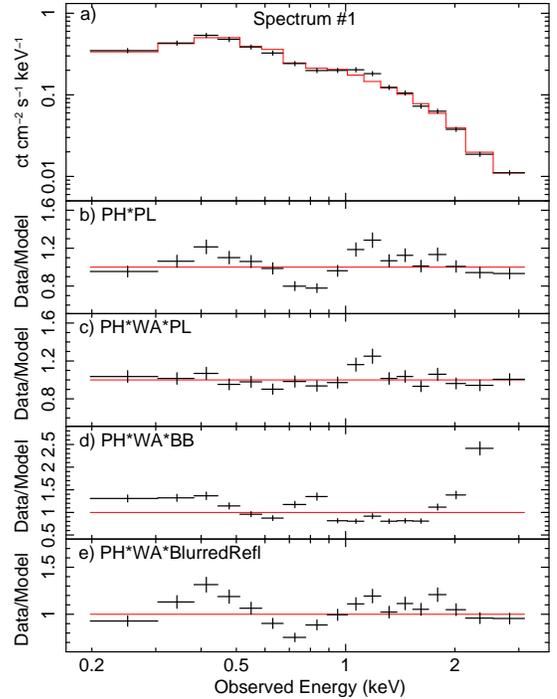}
\caption{\label{fig:rmsspec1_detail}
Spectral fits to the lowest-frequency \rms\ spectrum (\#1; timescale of 100 ks).
The top panel shows the data and best-fit model.
Panel b shows the data/model residuals to a power-law modifed by cold absorption (in excess of
the Galactic column).
Panel c shows the data/model residuals for the best-fit model,
which includes the low-ionization warm absorber.
In panel d, the power law has been replaced by a blackbody component; this yields a very poor fit.
In panel e, the continuum emission is modeled by blurred reflection
off an ionized medium; this also yields a very poor fit.}
\end{figure}

We started with spectrum \#1, re-plotted in Fig.~\ref{fig:rmsspec1_detail}.
We added a full-covering neutral absorber with \textsc{zphabs}; $\chi^2/dof$ fell to 85.2/15=5.7 for
$\Gamma\sim2.45$ and column density $N_{\rm H,cold} \sim 3\times10^{20}$ cm$^{-2}$, but data/model residuals (Fig.~\ref{fig:rmsspec1_detail}b)
were still poor.
Finally, we modeled a full-covering zone of ionized absorption
using the same \textsc{xstar} table as M09. Redshifts for all ionized absorbers
were frozen at the values from fits to the time-averaged RGS spectrum in M09; 
all abundances were frozen at solar values.
This fit yielded $\chi^2/dof=27.9/13=2.15$,
with best-fit parameters listed in Table~\ref{tab:rmsspec_table}, including $\Gamma=3.03\pm0.17$,
nearly consistent with that for the soft X-ray power-law in the time-averaged spectrum. 
Data/model residuals (Fig.~\ref{fig:rmsspec1_detail}c) are generally acceptable except for two points at 1.1--1.25 keV, but there are no clear
candidates for what causes this deviation. One might expect some Fe L edge absorption near 1.0--1.3 keV due
to lowly- or moderately-ionized gas, but only at the $\sim6\%$ level for the best-fit values of log($\xi$) and column density
$N_{\rm H,WA}$; allowing the Fe abundance relative to solar vary had no effect on the fit.

The \rms\ spectrum is not highly sensitive to the
higher-ionization (log($\xi$)$\sim2.9$)  absorber modeled in the EPIC and RGS energy spectra.
This absorber affects the continuum only at the $\sim$ few percent level. It also
introduces discrete absorption lines, the strongest due to H-like O and Ne and He-like Fe, 
but the \rms\ spectrum lacks the energy resolution to detect them. 

We tested if the continuum could also be modeled by thermal Comptonization, as per fits to the time-averaged
EPIC spectrum. We replaced the power law with a \textsc{compST} component.
The best-fit model (see Table \ref{tab:rmsspec_table}) yielded 
data/model residuals virtually identical to those for the power-law, as the Comptonized component
effectively resembles a power law, with a slight bend above $\sim2$ keV.
Best-fit parameters for optical depth $\tau$ and electron temperature
$k_{\rm B}T_{\rm e}$ are factors of roughly $\sim$2 lower and $\sim$3 higher, respectively, than in M09. However,
there exists some degeneracy between these two parameters, and they are anticorrelated
in the direction of the best-fit values of M09.

Next, we replaced the \textsc{compST} component with a blackbody component, as M09
were able to model the soft excess with a phenomenological blackbody with temperature
$k_{\rm B}T \sim 83$ eV. However, this yielded an extremely poor fit ($\chi^2/dof=595/13\sim46$ 
for $k_{\rm B}T \sim 200$ eV; data/model residuals in Fig.~\ref{fig:rmsspec1_detail}d), 
and we conclude that an absorbed blackbody is not able to describe the variable component.

Finally, we also tested models incorporating reflection from an ionized
medium, using \textsc{reflion} (Ross \& Fabian 2005), and tested both
unblurred and blurred emission, the latter using \textsc{kdblur}.
However, we did not achieve a satisfactory fit despite trying the fit
with a wide range of initial parameters and with all parameters except
outer radius of blurring left free; our best fit had $\chi^2_{\rm red}
\sim 12$ for a model with $\xi\sim 300$ erg cm s$^{-1}$, inner radius
of blurring $\sim 1-3 R_{\rm g}$ and an emissivity index $q$ near 0.
Adding the low-ionization warm absorber improved the fit quality only
slightly, with $\chi^2_{\rm red} \sim 9$ (best-fit parameters include
$\xi\sim400$ erg cm s$^{-1}$, inclination angle $\sim85$ degrees,
inner radius of blurring $\sim 1-2 R_{\rm g}$). The very strong
data/model residuals are plotted in Fig.~\ref{fig:rmsspec1_detail}e.
Any model with significantly lower values of $\xi$ had a spectral
shape above 1 keV that was much too flat to match the observed
spectrum.  Adding in neutral absorption via \textsc{zphabs} did not
significantly improve any of the above fits.

\subsection{High-frequency \rms\ spectra}
\begin{figure}
\psfig{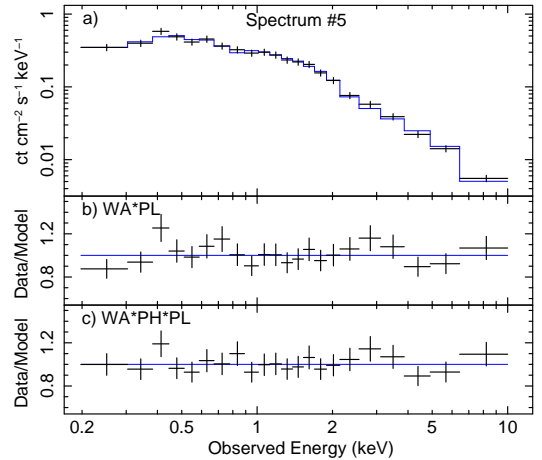}
\caption{\label{fig:rmsspec5_detail}
Spectral fits to the lowest-frequency \rms\ spectrum (\#5; timescale of 19.8 ks).
The top panel shows the data and best-fit model.
Panel b shows the data/model residuals to a power-law modified the low-ionization absorber 
(in addition to the Galactic column).
Panel c shows the data/model residuals for the best-fit model,
which includes both excess cold absorption and the low-ionization warm absorber.}
\end{figure}

We now turn to spectrum \#5, the higher-temporal frequency spectrum. 
We fit the 0.2--10 keV spectrum; the simple power-law model above yielded
$\chi^2/dof$=29.5/20=1.5 for $\Gamma\sim1.57$.
Data/model residuals (Fig.~\ref{fig:rms_spec_first}b) had a very similar shape to those for the low-frequency
fit with a power law, indicating that similar zones of absorption may be appropriate to model.
We tried modeling Compton reflection using the \textsc{pexrav} component,
and using values from M09, but this had no effect on the fit due to the
limited energy range of EPIC and is not discussed further. 
We added a warm absorber with log($\xi$)$\sim 0.6$; data/model residuals are plotted in 
Fig.~\ref{fig:rmsspec5_detail}b.
$\chi^2/dof$ fell to 16.7/18=0.93 i.e., already an acceptable fit. For completeness, however, and to improve residuals $<$0.4 keV, we
added a neutral absorber to the model with \textsc{phabs} and obtained our best-fit model; best-fit parameters
are listed in Table~\ref{tab:rmsspec_table} 
and data/model residuals are plotted in Fig.~\ref{fig:rmsspec5_detail}c.
$\Gamma$ is $1.87^{+0.15}_{-0.10}$, similar  
to that for the hard power-law component in the time-averaged spectrum.
We will demonstrate
in the discussion section that the hard spectral components in the \rms\ and time-averaged
spectra are in fact consistent with each other, i.e., we confirm the hard X-ray power-law as
the only variable component on these timescales.  

\begin{figure}
\psfig{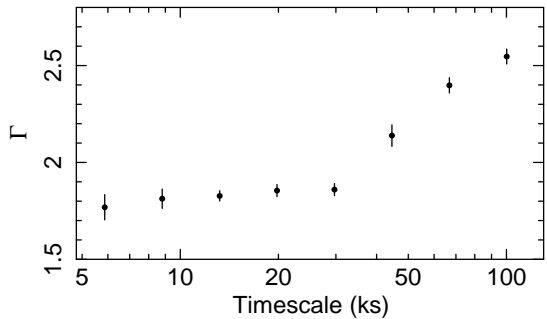}
\caption{\label{fig:Gamma_time} Best-fit values of photon index for each
\rms\ spectrum as a function of variability timescale, and after accounting for
line of sight absorption following M09. 
The spectrum of the variable component changes abruptly at a variability timescale of 30 ks, close to the break frequency in the power spectrum.}
\end{figure}

To track the spectral slope and thus the dominating spectral component as a function of 
timescale, we simultaneously fit all eight \rms\ spectra with a model consisting of
a power-law modified by the cold absorber and low-ionization warm absorber, with column densities each left free.
The value of log($\xi$) was fixed at 1.43.
We included for completeness the high-ionization absorber with log($\xi$) fixed at 2.89 and column density frozen at 
$1.5 \times 10^{21}$ cm$^{-2}$.
The resulting values of $\Gamma$ as a function of timescale are plotted in Fig.~\ref{fig:Gamma_time}.
A change in spectral slope occurs sharply at a variability timescale of 30 ks, close to the break frequency in the PDS.
We conclude that a soft spectral component alone is producing the long-term trend seen through the \xmm\ observation, 
while all the rapid fluctuations are produced by a hard power-law component with $\Gamma \sim 1.6-1.7$.

\begin{deluxetable*}{lccc}
\tablecaption{ Model Fits to Low- and High-Frequency RMS Spectra}
\tablecolumns{4}
\tablewidth{6in}
\startdata \hline
                          & Low Frequency                          & Low Frequency                 &  High Frequency \\
                          & (power law)                            & (\textsc{compST})             &   (power law)   \\   \hline
$\chi^2/dof$              & 27.9/13=2.15                           & 26.7/12=2.22                  & 10.9/17=0.64        \\        \hline
$\Gamma$                  & $3.03\pm0.17$                          &                               & $1.87^{+0.15}_{-0.10}$ \\    
$A_1$ (ph cm$^{-2}$ s$^{-1}$ keV$^{-1}$) & $3.60^{+0.65}_{-0.50} \times 10^{-4}$  &                & $5.0^{+1.3}_{-0.9} \times 10^{-4}$ \\  \hline
$k_{\rm B}T_{\rm e}$ (keV) &                                       & $1.0^{+8.0}_{-0.3}$              &  \\
$\tau$                    &                                       & $14^{+4}_{-13}$                     &  \\
Norm.\                    &                                       & $3.5^{+0.6}_{-0.5} \times 10^{-4}$  &  \\   \hline
$N_{\rm H}$ (neutral; $10^{20}$ cm$^{-2}$)   & $6^{+3}_{-2}$        &  $6\pm3$                      &  $<6.7$  \\
$N_{\rm H}$ (ionized; $10^{21}$ cm$^{-2}$)   & $2.0^{+0.6}_{-0.5}$  & $2.2^{+1.0}_{-0.6}$           & $3.0^{+7.0}_{-1.5}$  \\
log($\xi$, erg cm s$^{-1}$)  &  $0.85^{+0.25}_{-0.35}$              & $1.00^{+0.25}_{-0.40}$        & $1.35\pm0.60$ \vspace{+0.3cm}   
\enddata
\tablecomments{Model parameters for best fits to low- and high-frequency \rms\ spectra (\#1 and \#5, respectively).
The reader is referred to the \textsc{Xspec} 
help manual for the units of the normalization of the \textsc{compST} component.
$A_1$ is the normalization of the power law at 1 keV.
\label{tab:rmsspec_table}}
\end{deluxetable*}

\subsection{Variable cold absorption}

We investigated if the observed smooth flux change affecting mainly the
soft energy bands could be due to varying cold absorption, for example
by a cloud crossing the line of sight on timescales of 100 ks. In this
case, considering that the hard power law is not varying strongly on timescales
of about 50 ks or longer, the low-frequency \rms\ spectrum should have
the same shape as the difference between the spectra with initial and
final absorption values. We explored this possibility by constructing
difference spectra of variable cold absorption and comparing their shapes
to that of the \rms\ spectrum of slow fluctuations.

\begin{figure}
\psfig{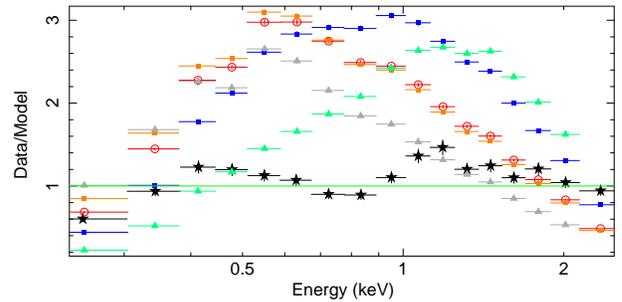}
\caption{\label{full_covering} Comparison of the low-frequency \rms\ spectrum, denoted by black stars, 
to the difference spectra produced by variable full-covering cold absorption.  
The red open circles correspond to a change in \nh\ from 3 to 10 $\times 10^{20}$ cm$^{-2}$, encompassing
the best-fit time-averaged value; orange and blue squares represent even larger changes in \nh\ still covering the time-averaged
values of $N_{\rm H}$. 
Triangles show cases of more extreme changes in $N_{\rm H}$, 
with both initial and final values of \nh\ above (green) and below (grey) the time-averaged value. 
All spectra have been divided by a power-law model of equal slope to highlight the differences in their shapes. 
Evidently, varying the column of a full-covering absorber cannot explain the shape of the low-frequency \rms\ spectrum and 
therefore cannot be the cause of the slow trend in the soft band light curves.}
\end{figure}

The difference spectra were constructed by taking the best-fitting
model to the time-averaged spectrum of M09, varying the value of 
$N_{\rm H,cold}$ (henceforth simply $N_{\rm H}$, with units of cm$^{-2}$ for all values discussed) 
of the full-covering, cold absorber around its best-fitting value and taking the
difference of model spectra with different values of $N_{\rm H}$. If the observed slow/soft
variability is produced only by a change in \nh\ then the \rms\
spectrum will have the same shape as the difference spectrum, although
the normalization can be different. We fitted a simple power-law model
to the difference and \rms\ spectra tying together their photon
indices and leaving their normalizations free. The ratio of these
spectra to the best fitting power-law model highlights the difference
between the spectra while correcting for their different
normalizations. Since we are only interested in the comparison of the
shape of the low-frequency \rms\ spectrum to the difference spectra
produced by variable cold absorption, these ratio plots are
sufficiently informative and a sample of difference spectra is shown
in Fig.~\ref{full_covering}.

We started by calculating difference spectra with initial and final
values of \nh\ of $3 \times 10^{20}$ and $10 \times 10^{20}$,
respectively. These straddle the best-fit time-averaged value of 
$7 \times 10^{20}$ and produce the observed 
factor of two increase observed in the flux of the 0.2--1 keV light
curve. We also tested different values of \nh\ covering
a wider range but still straddling the time-averaged value, such as the
1--10 $\times 10^{20}$ and 3--30 $\times 10^{20}$ difference spectra
plotted in Fig.~\ref{full_covering} in orange and blue triangles,
respectively.

All difference spectra investigated have a characteristic peak that
deviates from a simple powerlaw significantly more than the observed
\rms\ spectrum does. We therefore conclude that the slow trend seen in
the soft light curve cannot be produced by a change in the \nh\ of a
full-covering cold absorber.

\begin{figure}
\psfig{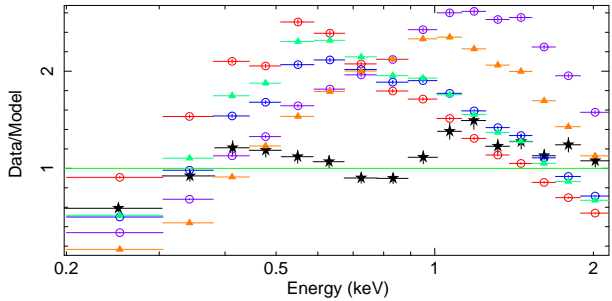}
\caption{\label{pcfigure}Comparison of the low-frequency \rms spectrum, in black stars, to the difference spectra produced by variable covering fractions, plotted in circles, and variable \nh\ for a 50 \% covering fraction, plotted in triangles. The \nh\ values used for the variable covering fraction spectra were 5, 20 and 80 $\times 10^{20}$, plotted in red, blue and purple circles. For the constant 50\% covering fraction models we used \nh\ changing from 3 to 10 $\times 10^{20}$(green triangles) and 10 to 30$\times 10^{20}$ (orange triangles). A varying covering fraction or variable \nh\ of a partial coverer do not produce difference spectra that can explain the low-frequency \rms spectrum.}
\end{figure}

Changes in partial covering are also unable to reproduce the shape of
the \rms\ spectrum. We first constructed difference
spectra where the initial and final spectra were subject to partial covering
of cold gas (using \textsc{pcfabs} in \textsc{Xspec}),
where the covering fraction varied but the column density $N_{\rm H,PC}$ did not.
The shape of these difference spectra only
depends on the value of $N_{\rm H,PC}$, not on the initial and
final covering fractions.  Only the normalization of these difference
spectra is proportional to the difference in covering fraction. Since
the normalization is modeled out by dividing the \rms\ and difference
spectra by powerlaw models with their best-fitting normalization,
variations in only one parameter need to be explored. The difference
spectra of these variable-covering fraction models are plotted with circles in
Fig.~\ref{pcfigure} for values of $N_{\rm H,PC} = 5$, 20 and 80 $\times 10^{20}$,
plotted in red, blue, and purple circles, respectively. The value of $N_{\rm H,PC}$ 
determines the position of the peak in the difference spectra compared to a
simple power law, but the shape remains relatively constant, increasing
much more steeply than the \rms\ spectrum below the peak and also
decreasing much more steeply after. The final scenario we tested 
was a partial coverer whose column density $N_{\rm H,PC}$ varies
while the covering fraction remains the same. The resulting
difference spectra are qualitatively similar to those described above and a couple
of examples are plotted in Fig.~\ref{pcfigure} in triangles, for a
constant covering fraction of 50\% and values of $N_{\rm H,PC}$ changing from 3 to $10 \times
10^{20}$ (green triangles) and 10 to $30 \times 10^{20}$ (orange
triangles). We conclude that a partial coverer 
with either varying covering fraction or variable column density
does not produce difference spectra that can
explain the observed low-frequency \rms\ spectrum or the smooth increase
in soft band flux.

\section{Interpretation of the variability spectra}
\label{interpretation}
We have used coherence and Fourier-resolved spectral analysis to
deconvolve two independently-varying continuum components: a 
soft spectral component consistent with a power law of Photon Index
$\Gamma=3.03\pm0.17$ that varies on timescales of 30--100 ks, and
a hard power-law component of $\Gamma
=1.87\pm0.15$. The hard power law component accounts for 
variability across the entire energy range on all shorter timescales. In
this section we will explore different possible scenarios to explain these components
and their behavior during this observation and we compare our results to previous studies.

\subsection{Slowly-varying soft component}

The long-term trend, evident in the 0.2--1 keV light curve of
Fig.~\ref{fig:lcs}, dominates the variance of the lowest frequency
\rms\ spectra. The spectral shape corresponding to this fluctuation is
very soft, and can be fitted with a power law of $\Gamma=3.03\pm0.17$
plus neutral and warm absorption as demonstrated in
Fig.~\ref{fig:rmsspec1_detail}.

The low frequency \rms\ spectrum is reminiscent of the soft excess
component in the spectral fit to the time-averaged spectrum of \ngc\
by \citet{markowitz_3227}. The
\rms\ spectrum can also be modeled well by a \textsc{CompST} component with
parameters roughly consistent to those in the time-averaged spectrum, also affected by warm absorption. The
consistency between warm absorption parameters on the \rms\ and
time-averaged spectra argues for a \emph{constant} warm absorber
acting on a soft excess of variable normalization. 
Thermal Comptonization models have been successfully 
used to model soft excesses previously
in a few Seyfert spectra \citep{jin09,middleton09}.
We also demonstrated that a single-temperature blackbody, despite
providing a plausible phenomenological description of the soft excess
in the time-averaged spectrum, cannot yield a satisfactory
fit to the low-frequency \rms\ spectrum.

The soft excess varies independently of the hard power law;
additional examples of this behavior in other Seyferts' X-ray 
spectra are given in e.g.\ \citet{rivers12,markowitz_3227} (although cases where the soft excess varies with the harder powerlaw have also been observed, e.g.  1H0707-495 \citep{fabian09},IRAS 13224Ð
3809\citep{ponti10}). The independent variability is
evident in the low-frequency \rms\ spectrum, where only the soft
excess appears, and in the higher frequency \rms\ spectra, where the
powerlaw is present and the soft excess disappears.
The power spectrum of the soft excess cannot be studied in detail
since this component only exhibits a slow trend throughout the
observation. The long-term \rxte\ light curves cannot provide additional
information on the long term behavior of the soft excess since
they only track energies above 2 keV. 

The optical/UV light curves presented in \citet{markowitz_3227} show a
steady increase in flux by $\sim10\%$ throughout the observation, qualitatively
similar to the increase seen in the soft X-rays (factor of $\sim2$ 
increase throughout the observation). { This behavior is 
consistent with a connected optical/UV/soft X-ray trend. It is not possible, however, to establish the significance of this correlation since the timescale of the trend is similar or longer than the duration of the observation, and it is not unlikely that uncorrelated trends on these timescales would look similar by chance. \footnote{In principle, the same argument applies to the slow variations of the soft X-ray energy bins, they could be uncorrelated narrow emission features that varied in the same way during the observation. Although such a spectral decomposition is not usual and the chance correlation of several bands is less likely, it is a possibility that cannot be ruled out.} }
A similar optical--soft X-ray correlation has been observed in
Mkn~509 \citep{Mehdipour}. 
These authors find a soft X-ray excess and hard
X-ray power law that each vary independently over timescales of a few
days, but the addition of a simultaneous UV light curve shows that the soft
excess does correlate well with the variability at lower energies. 
Mehdipour et al.\ interpret this soft excess as Comptonized thermal disc
photons by a corona of optical depth $\tau\sim17$ and electron temperature
$k_{\rm B}T_{\rm e}\sim0.2$ keV in the vicinity of the accretion disc. In our case, the
variable (time-averaged) soft excess can be modeled as a Comptonized component with
$\tau\sim14$ (24) and $k_{\rm B}T_{\rm e} \sim 1.0$ (0.35) keV; importantly,
the soft excess also displays uncorrelated variability to the hard power law and a closer connection
to the optical/UV behavior. Therefore, a similar interpretation is
warranted for the soft excess in \ngc . This interpretation of the
origin of the soft excess emission was previously discussed in
\citet{markowitz_3227} together with other observations of variable
soft excesses.

As discussed in \citep{markowitz_3227}, the soft excess in \ngc\ doubles its 
flux on a timescale of 50 ks. This duration 
corresponds to a light travel time of $\sim$1000 $R_{\rm g}$, where
$R_{\rm g} \equiv GM_{\rm BH}/c^2$, or $\sim$10 times the orbital timescale at 6$R_{\rm g}$. 
This timescale might be too short
for accretion rate fluctuations produced by viscous processes in the
geometrically thin accretion disc, but might originate in a
geometrically thick corona.

Finally, the low frequency \rms\ spectrum drops steeply with energy up
to $\sim3$ keV but it begins to rise again above this energy. The lack of
correlation between low and high energy slow fluctuations argues
against a single component for the whole low-frequency \rms\
spectrum. It is tempting to identify the higher-energy portion of
the slowly-varying component
with a Compton reflection hump varying on timescales of $>$30 ks
(i.e., 1000 $R_{\rm g}$/c) and not on shorter timescales. \citet{rivers11}
fit the time-averaged \textit{RXTE} spectrum for this source using
and measured only a very weak
reflection hump (a value of reflection strength $R=0.3\pm0.1$  
using the \textsc{pexrav} model component), which would only account for approximately 3\% of the
total flux in the 6--10 keV band. Therefore, if the Compton hump were
varying as much as the soft excess at low frequencies ($F_{\rm var}=6.7\%$), we
would only expect a value of $F_{\rm var}=6.7\%/33=0.2\%$ in the 6.4--10 keV band, in
contrast to the value observed, $F_{\rm var}=1.7\%$. The Compton hump therefore could
produce only a small fraction of the slow variations in the hard
bands, which are more likely produced by the power-law component.

{ Finally, we note that blurred, ionized reflection as a source for the soft excess did not fit well the spectrum of this source below 3 keV, so the low-frequency \rms\ spectrum does not correspond to a disc reflection component varying in normalization. In the disc reflection scenario, the high amplitude variation observed in the soft excess could be produced by changes in the ionization parameter instead, affecting mainly the line emission, below 2 keV. The powerlaw component did not vary on these long timescales during this observation, so the large change in ionization state could not have been produced by a similar change in illumination, unless it were responding to earlier fluctuations. The delay in the response of the disc reflection would have to be longer than the observation length, i.e. 100 ks, which equals the light travel time across several thousand gravitational radii in this AGN, making this scenario unlikely.    }

\subsection{Rapidly-varying power-law component}

{ The \rms\ spectrum of the rapid fluctuations in this observation has a power-law shape over the 0.2--10 keV range,  affected by constant cold and warm absorption. The parameters of the absorbers are consistent with those measured in the time-averaged spectrum (M09) and in the slowly-varying soft excess. Similar \rms\ spectra have been observed in other Seyfert galaxies, e.g. MCG--6-30-15 \citep{Papadakis2005} and Mkn~766 \citep{arevalomkn766}, where the variability spectrum is approximately a power law with a broad absorption feature around 0.7--1 keV. This common feature in \rms\ spectra of Seyfert galaxies with soft excesses indicates that constant warm absorption acting on the power law component is present in all these sources. As shown above, the spectrum of \ngc\ has at least two separate continuum components, the soft-excess discussed in the previous section and a hard powerlaw. The warm absorption appears around the energies where these components intersect, modifying the time-average spectrum further.  }

The spectral variability of NGC~3227 as observed in
the long term \rxte\ monitoring data presented by \citet{uttleymchardy05}
indicated that pivoting of the hard power-law is the main source of
flux variability in the 2--15 keV band. The pivot point was calculated
to be at a few hundred keV. For energies far below the pivot point,
the fluctuations resemble simple changes in normalization, but where
the amplitude of variations increases with decreasing energy. The
\rms\ spectrum expected from such a pivoting powerlaw is also a
powerlaw but with slightly steeper slope, to account for the larger
fluctuations at lower energies.

With this \xmm\ observation we can study the spectral shape of this variable component down to lower energies. The coherence analysis for rapid (less than 30 ks) fluctuations indicates that the \rms\ spectrum corresponds to a single spectral component down to 0.2 keV. Its shape is a hard power law of $\Gamma \sim 1.75-1.85$.  
\citet{markowitz_3227} has fitted the time-averaged energy spectrum with several components including a power law to account for most of the hard band flux. The fitted slope in their `SXPL" broadband model fit
was $\Gamma=1.57 \pm 0.02$, flatter than the \rms\ spectral slope and consistent with the pivoting scenario.

\citet{uttleymchardy05} find that the long term fractional variability (i.e. \rms /count rate) of \ngc\ in the 3--5 keV band is 50\% while in the 7--15 keV band it is 38\%. For a time-averaged photon index $\Gamma$ of 1.6, this dependence of fractional variability on energy band corresponds to an \rms\ spectrum with $\Gamma=1.9$, similar to that observed in the \xmm\ observations for the 0.2--10 keV band. The consistency between \rms\ slopes in the hard band for long term \rxte\ and short term \xmm\ light curves argues for a single hard power-law component producing all the observed variability above 2 keV, while the low-frequency \rms\ spectrum with $\Gamma \sim 3$ in this \xmm\ observation is produced by a separate spectral component.  

\citet{Papadakis2007} constructed \rms\ spectra for six Seyfert galaxies and compared their slopes to the stellar-mass black hole X-ray binary (BHXRB) 
Cyg X-1 in its hard and soft states \citep{revnivtsev99,gilfanov00}. They fit the spectra above 3 keV to characterize the slope of the variable hard powerlaw so we can only compare their results to the high-frequency \rms\ spectra of our \xmm\ observation of \ngc . They find the six AGN \rms\ spectral slopes to lie between the hard and soft states of Cyg X-1 and become flatter with increasing frequency at a rate of $\Delta \Gamma \propto f^{-0.25}$. The comparison between objects of different black hole mass is facilitated by scaling the variability frequencies to the Keplerian frequency $f_{\rm K}$ of each object at 3 Schwarzschild radii. Using a black hole mass for \ngc\ of $10^{6.88}$ \Msun\ from \citet{Denney10}, our measurements for \ngc\ fall slightly below the relation of \citet{Papadakis2007}; at $f/f_{\rm K}=0.6$ we measure a slope of 1.76 vs.\ 1.95 measured for other AGN and at $f/f_{\rm K}=0.12$ we find 1.87 vs.\ 2.3. The \rms\ slopes we measure do not depend as strongly on frequency as the composite of the sample of the other six AGN, with a $\Delta \Gamma \propto f^{-0.03}$ but the trend is at least in the same direction.

The accretion rate of \ngc\ relative to Eddington was estimated by \citet{vasudevan09} from
simultaneous \xmm\ X-ray and optical/UV observations to be
$L_{\rm bol}/L_{\rm Edd}=0.45 - 1.5 \times 10^{-3}$. However, 
the black hole mass estimate from reverberation-mapping $M_{\rm BH}$ has since
been revised from $10^{7.63}$ \Msun\ \citep{revmap} to $10^{6.88}$
\Msun\ \citep{Denney10}, yielding $L_{\rm bol}/L_{\rm Edd}=2.5 - 8.5 \times 10^{-3}$.
Using \textit{Swift}-BAT and \textit{IRAS} observations (and the updated value of 
$M_{\rm BH}$), \citet{vasudevan10} derive
$L_{\rm bol}/L_{\rm Edd}=2.0-3.9 \times 10^{-2}$. Even this corrected
estimate of the accretion rate is significantly lower than the average
of the six AGN in the \citet{Papadakis2007} sample of $L_{\rm bol}/L_{\rm Edd}=0.65$
\citep[and reference therin]{uttleymchardy05}. The high-frequency \rms\ spectra of
\ngc\ are flatter than those found for other AGN. These values of $\Gamma$ do, however,
fall between the values of the \rms\ spectral slopes of Cyg X-1 in soft and hard
states, but closer to the hard state values than the other AGN
investigated. Therefore, our results combined with those of
\citet{Papadakis2007} are consistent with a picture where the \rms\
spectra flatten with decreasing accretion rate, both in AGN and in the
BHXRB Cyg X-1.

As a final note, we recall the position of \ngc\ within the plane
describing the dependence of X-ray PDS break timescale $T_{\rm b}
\equiv 1/f_{\rm b}$ on both $M_{\rm BH}$ and $L_{\rm Bol}$, applicable
to Seyfert AGN and BH XRBs simultaneously, and empirically quantified
by McHardy et al.\ (2006).  Using the updated value for the black hole
mass estimate and using the values of $L_{\rm Bol}$ reported by
Vasudevan et al.\ (2010), $10^{43.2-43.5}$ erg s$^{-1}$, the empirical
relation of McHardy et al.\ (2006) (using the best-fit values for the
coefficients and ignoring the coefficient uncertainties for the
moment) yields predictions for $T_{\rm b}$ of 1.04--2.05 d, completely
consistent with our measured value of $T_{\rm b} =
1.01^{+2.39}_{-0.50}$ d.  For completeness, using values of $L_{\rm
Bol}$ from Woo \& Urry (2002; $10^{43.9}$ erg s$^{-1}$), as was done
by UM05, or Vasudevan \& Fabian (2009) ($10^{42.4-42.9}$ erg s$^{-1}$)
instead yields predictions for $T_{\rm b}$ of 0.42 d or 4.1--12.5 d,
respectively.

\section{Conclusions}
\label{conclusion}

We studied the energy- and temporal-frequency-dependent behavior of the X-ray
continuum emission components in the Seyfert 1.5 AGN NGC~3227
using a 100 ks long-look obtained with \textit{XMM-Newton} in 2006. 

We revisited the broadband 2--10 keV PDS of this source, first published by UM05, by
combining the PDS derived from the \xmm\ light curve with PDSs derived from multi-timescale
\textit{RXTE} monitoring. We applied a bending power-law model, and obtained a best-fit bend frequency
of $1.15^{+0.76}_{-0.81}\times 10^{-5}$ Hz ($1.01^{+2.39}_{-0.40}$ d), 
consistent with the results of UM05 and consistent with the empirical
relation between X-ray black hole mass, and X-ray luminosity
of McHardy et al.\ (2006) considering updated measurements of the black hole mass.

The \xmm\ observation coincided with a period of low variability amplitude above 2 keV on timescales of 
50--100 ks. This chance occurrence allowed us to identify a separate soft component varying 
slowly in normalization, and independently of variations in the coronal hard X-ray power-law component.
More specifically, the timescale-resolved \rms\ spectra together with the coherence
analysis allowed us to isolate and quantify the variable components:
the steep soft excess varying coherently from 0.2 up to 2 keV
and dominating the variability on timescales longer than 30 ks,
and the harder power-law component, varying coherently from 0.2--10 keV
and dominating the variability on timescales shorter than 30 ks.

Both components are affected to the same extent as the time-averaged spectrum by 
absorption by full-covering, neutral and ionized gas along the line of sight.
The cold and warm absorbers are constant on timescales covered by this 
observation and the variability is intrinsic to the emitting regions.

The variable soft excess component can be 
well-described by a steep ($\Gamma \sim 3$) power-law. It can also
be described by a model of thermal Comptonization, 
in which optical/UV seed photons are upscattered by a distribution of hot electrons with optical depth 
$\sim 14-24$ and electron temperature $k_{\rm B}T \sim 0.3-1$ keV. 
A blackbody component, despite being an acceptable phenomenological fit to the soft excess in the
time-averaged spectrum, is ruled out by the \rms\ spectra, as is a blurred ionized reflection component.

The \rms\ spectrum of the hard power law has a slope of $\Gamma \sim 1.8$, similar to but steeper than the power law fitted to the time-averaged spectrum. This steepening is consistent with the pivoting mechanism proposed to model the hard X-ray variability of this object from long-term light curves in the 2--15 keV band \citep{uttleymchardy05}. The hard power-law variability is not well-correlated with either the soft excess or optical/UV fluctuations on the time-scales studied, 5--100 ks, which supports the division of the X-ray spectrum in at least two emission components. { The rapid fluctuations that characterize the powerlaw component show hard lags at a level of 2\% of the variability timescale, in line with similar findings of powerlaw variability in other AGN.} A longer monitoring campaign would be needed to test whether the optical/soft X-ray and hard X-ray bands are correlated on long timescales, as is often the case in other Seyfert galaxies.

\section*{Acknowledgements}
This work is based on observations obtained with \xmm , an ESA science mission with instruments and contributions directly funded by ESA Member States and NASA and has made use of HEASARC online services, supported by NASA/GSFC, the NASA/IPAC Extragalactic Database, operated by JPL/California Institute of Technology under contract with NASA, and the NIST Atomic Spectra Database. PA acknowledges financial support from Fondecyt grant 11100449 and Anillo ACT1101.

\bibliographystyle{apj}

\label{lastpage}

\end{document}